\documentclass[10pt,journal]{IEEEtran}
\IEEEoverridecommandlockouts
\usepackage{cite}
\usepackage{amsmath,amssymb,amsfonts}
\usepackage{algorithm}
\usepackage{algpseudocode}
\usepackage{graphicx}
\usepackage{textcomp}
\usepackage{enumitem}
\usepackage{xcolor}
\usepackage{color}
\usepackage{siunitx}
\usepackage{footnote}
\usepackage{float}
\usepackage{threeparttable}
\usepackage{makecell}
\usepackage{subfig}
\usepackage{subfloat}
\usepackage{booktabs}
\usepackage{bm}
\usepackage{multirow}
\usepackage{bigstrut}
\usepackage{mathrsfs}

\makeatletter

\newcommand{\Rmnum}[1]{\expandafter\@slowromancap\romannumeral #1@}

\makeatother

\graphicspath{{figure/}}
\def\BibTeX{{\rm B\kern-.05em{\sc i\kern-.025em b}\kern-.08em
    T\kern-.1667em\lower.7ex\hbox{E}\kern-.125emX}}
\begin{document}
    \captionsetup[figure]{name={Fig.},labelsep=period}

    \title{Spatio-Temporal-Frequency Graph Attention Convolutional Network for Aircraft Recognition Based on Heterogeneous Radar Network\\}

    \author{Han Meng, \IEEEmembership{Student Member, IEEE,}
	Yuexing Peng, \IEEEmembership{Member, IEEE,}
	Wenbo Wang, \IEEEmembership{Member, IEEE,}
	Peng Cheng, \IEEEmembership{Senior Member, IEEE,}
	Yonghui Li, \IEEEmembership{Fellow, IEEE,}
	and Wei Xiang, \IEEEmembership{Senior Member, IEEE}\\
	
	    \thanks{H. Meng, Y. Peng and W. Wang are with the School of Information and Communication Engineering, Beijing University of Posts and
		Telecommunications, Beijing, 100876, China. e-mail: \{menghan, yxpeng,	 wbwang\}@bupt.edu.cn}% <-this % stops a space
	    \thanks{P. Cheng is with the Department of Computer Science and Information Technology, La Trobe University, Australia, and also with
		the University of Sydney, Australia (e-mail: p.cheng@latrobe.edu.au;
		peng.cheng@sydney.edu.au).}% <-this % stops a space	
	    \thanks{Y. Li is with the School of Electrical and Information Engineering, The
		University of Sydney, NSW 2006, Australia. email: yonghui.li@sydney.edu.au}% <-this % stops a space
	    \thanks{W. Xiang is with the School of Engineering \& Math Sciences, La Trobe University, Australia (e-mail: w.xiang@latrobe.edu.au).}% <-this % stops a space	 
        \thanks{This work is supported by NSFC under grant 62071063. Corresponding authors: yxpeng@bupt.edu.cn and wbwang@bupt.edu.cn}}
	    %\thanks{Manuscript received April 19, 2005; revised August 26, 2015.}

% The paper headers
    \markboth{Journal of \LaTeX\ Class Files}%
    {Shell \MakeLowercase{\textit{et al.}}: Bare Demo of IEEEtran.cls for IEEE Journals}
% The only time the second header will appear is for the odd numbered pages
% after the title page when using the twoside option.
%
% *** Note that you probably will NOT want to include the author's ***
% *** name in the headers of peer review papers.                   ***
% You can use \ifCLASSOPTIONpeerreview for conditional compilation here if
% you desire.

\maketitle
\begin{abstract}
This paper proposes a knowledge-and-data-driven graph neural network-based collaboration learning model for reliable aircraft recognition in a heterogeneous radar network. The aircraft recognizability analysis shows that: (1) the semantic feature of an aircraft is motion patterns driven by the kinetic characteristics, and (2) the grammatical features contained in the radar cross-section (RCS) signals present spatial-temporal-frequency (STF) diversity decided by both the electromagnetic radiation shape and motion pattern of the aircraft. Then a STF graph attention convolutional network (STFGACN) is developed to distill semantic features from the RCS signals received by the heterogeneous radar network. Extensive experiment results verify that the STFGACN outperforms the baseline methods in terms of detection accuracy, and ablation experiments are carried out to further show that the expansion of the information dimension can gain considerable benefits to perform robustly in the low signal-to-noise ratio region.
    \end{abstract}


    \section{Introduction}
    \IEEEPARstart{A}{IRCRAFT} is widely used in both military and civilian applications due to high mobility, long range, and flexible deployment. However, with the rapidly increasing capabilities of aircraft on defense penetration and low detectability in complex electromagnetic environments, airspace security and privacy become a very challenging task. As a countermeasure, aircraft recognition aims to classify target aircrafts by radar signals, and has attracted worldwide attention in recent years, where intensive in-depth studies on radar-based aircraft recognition have become a research focus \cite{zyweck1996radar,bhanu1997guest,zhang2012multi,molchanov2014classification,zhang2010time,yang2006target,zhang2019rcs,du2007radar,dongqing2016radar,kong2018automatic,ZhangMuqing2019the,guo2020variational,du2005radar,wang2011radar,kang2018micro,sun2016joint}.
    \par
     Radar based-aircraft recognition entails extracting characteristics from echo signals and then classifying the target aircraft. The echo signal of the target aircraft contains rich semantic information, such as velocity, shape, size, and acceleration. Under the umbrella of low detectability technology, it now becomes very difficult to detect characteristic parameters such as the electromagnetic radiation sharp information from weak radar echo signals reliably. Therefore, it is imperative to develop effective techniques to extract useful information from radar echo signals reliably and accurately.
    \par
    The commonly used radar echo signals can be roughly classified into three categories, i.e., 1) the radar cross-section (RCS) signal, 2) the high-resolution range profile (HRRP), and 3) the inverse synthetic aperture radar (ISAR) imaging. Although the HRRP \cite{guo2020variational,du2005radar,wang2011radar} provides precise information with a higher spatial and temporal resolution than the RCS signal, it requires high-resolution hardware and ultra-wideband radar. On the other hand, the ISAR \cite{kang2018micro,sun2016joint} mainly reflects the motion information, and is usually obtained from the airborne radar towards ground targets. Due to its easy availability and containing rich information of the target, the RCS signal is adopted in this work.
    \par
    Single radar-based aircraft recognition methods can be roughly categorized into statistical feature-based methods and end-to-end deep learning ones. Statistical methods attempt to detect some key feature parameters, such as the Doppler frequency \cite{molchanov2014classification,zhang2010time,yang2006target} or higher-order cumulant \cite{zhang2019rcs}, and then classify aircrafts using machine learning models. However, these methods require prior knowledge of the target aircraft which may not be available, and their performance degrades severely in the low signal-to-noise ratio (SNR) region. On the other hand, deep learning-based target recognition methods can automatically extract semantic features from abundant samples with limited direct human intervention \cite{kong2018automatic,guo2020variational,ZhangMuqing2019the}. For example, the methods in \cite{kong2018automatic} and \cite{ZhangMuqing2019the} process the low probability of intercept (LPI) radar signals for classification by convolutional neural network (CNN) and combining support vector machine (SVM) and autoencoder, respectively. In \cite{guo2020variational}, a deep generative recurrent network is developed to improve classification accuracy. Although superior recognition accuracy is achieved, the data-driven deep learning methods depend heavily on both the quantity and quality of the manually labeled data, whose performance inevitably deteriorates at low SNRs. Moreover, the data-driven methods, lacking recognizable mechanics analysis and relying only on grammatical feature learning from RCS signals, are unable to fully exploit semantic features. Therefore, these methods cannot maximize the performance of aircraft recognition.
    \par
    As a viable alternative, a homogeneous radar network \cite{chong2003sensor} consists of several spatially distributed radars of the same carrier frequency. These radars work collaboratively to receive spatially anisotropic scatter response waves simultaneously, which can greatly enhance the capability of aircraft recognition by means of temporal-spatial information fusion. Aiming for the low SNR region, the authors in \cite{859880} enhance the ability of weak signal detection through a quantitative detection fusion method. The authors in \cite{2012Distributed} propose a relaxed Chebyshev center covariance intersection algorithm to fuse local estimates when the cross-correlation of local estimation errors is unavailable. The distributed radar network multi-frame detection method
    is proposed in \cite{2018Multi}, which outperforms the single radar multi-frame detection method. The aforementioned methods extract the spatial-temporal information of the signals through signal-level fusion or decision fusion. However, there is still plenty of room for improving the detection performance, because these methods cannot extract grammatical and semantic information simultaneously. Besides, signal-level fusion methods require very high calibration accuracy between different radar information, so they are only suitable for homogeneous radar networks.
    \par
    When the collaborative radars in the radar network have different carrier frequencies and bandwidths, the original spatio-temporal domain is transformed into the much broader temporal-spatial-frequency domain, resulting in a heterogeneous radar network that can further enhance the detection performance. In this paper, we propose a framework to effectively distill and fuse temporal-spatial-frequency information in the heterogeneous radar network. Beyond signal-level fusion or decision fusion, we propose a model referred to as the spatial-temporal-frequency graph attention convolutional network (STFGACN), which is the first to exploit both grammatical and semantic features of the heterogeneous radar network. Based on the recognizability analysis of aircraft, the proposed STFGACN model maps the diverse grammatical features in the form of variation patterns of the RCS signal to the semantic feature determined by the unique movement pattern of the target aircraft. This greatly enhances the recognition performance in the low SNR region through fusing the diverse grammatical features in the temporal-spatial-frequency domain.
    \par
    The main contributions of this paper are highlighted as follows.
    \begin{itemize}
    	\item Based on the heterogeneous radar network, a network learning framework is developed to learn the mapping from semantic features determined by the unique movement pattern of the aircraft to the diverse grammatical features via temporal-spatial-frequency variation patterns of the RCS.
    	\item A novel STFGACN model is designed to distill semantic features and reliably detect aircraft by fusing diverse grammatical features in the form of information expansion from the temporal-spatial domain to the temporal-spatial-frequency domain. To the best of the authors' knowledge, this model is the first work to distill temporal-spatial-frequency information by means of the graph neural network (GNN) for aircraft recognition.
    	\item Extensive experiments are carried out to verify the superiority of the proposed model on the reliability of aircraft recognition, especially in the low SNR region.
    \end{itemize}
    \par
    The rest of the paper is organized as follows. Section \Rmnum{2} introduces the related work on the GNN. Section \Rmnum{3} formulates the aircraft recognition problem. Section \Rmnum{4} introduces the proposed framework and designed model. The performance evaluation is shown in Section \Rmnum{5}, followed by concluding remarks in Section \Rmnum{6}.
    \\ \hspace*{\fill}\\
    \emph{Notation}: $n$, $\bm{n}$, $\bm{N}$, and $\mathbb{N}$ represent a variable, vector, matrix, and set, respectively.

    \section{Related Work}
    The proposed STFGACN model is based on the recent proposed GNN \cite{kipf2016semi}, which works in non-Euclidean space and is a powerful technique to handle the real-world data featured by arbitrary graphical structures. According to the type of information extracted and the structure of the model, GNN can be roughly divided into the following three categories:
    \begin{itemize}
    	\item \textbf{Spatial-temporal propagation models:} Combining graph structure and time series features in the dynamic propagation process, GNN is proposed to model the spatial-temporal propagation of the disease state \cite{scarselli2009graph}. Based on the original GNN, many enhanced models are developed, e.g., cross-location attention based graph neural network (Cola-GNN) \cite{deng2020cola}, and spatio-temporal graph neural networks (STGNN) \cite{kapoor2020examining}.
    	\item \textbf{Spatial-temporal similarity-based models:} Based on the similarity among sub-nets in the spatial-temporal domain, graph convolutional network (GCN) based models are employed to distill spatial-temporal features for traffic forecasting \cite{yu2017spatio,chen2020multi}, weather forecasting \cite{seo2018automatically}, skeleton-based action recognition \cite{yan2018spatial}, and many other tasks \cite{bruna2013spectral, defferrard2016convolutional}.
    	\item \textbf{Semantic context extraction models:} By extracting syntax-based grammatical rules and context-based semantic of language, GNN models are designed to extract complex interactions between words or sentences in natural language processing \cite{yao2019graph,sui2019leverage,liu2020tensor}.
    \end{itemize}
    \par
    The differences between our model and the aforementioned models are summarized as follows. 1) Our model distills the semantic features depending on the aircraft movement pattern, while spatial-temporal propagation models model the grammatical representation in the form of state diffusion. 2) Spatial-temporal similarity-based models extract semantic features based on the similarity of both the temporal pattern of time-series and the spatial structure (e.g., the road network in traffic forecasting, geography in weather forecasting, and skeleton structure in skeleton recognition). By contrast, our model focuses on the identical semantic features determined by the movement pattern, rather than the spatial-temporal similarity. Moreover, the information dimension extension of the frequency domain represents a challenging task, which introduces a significantly increased complexity to handle the grammatical variation and diversity and can present more information for aircraft recognition through feature distillation. 3) Semantic context extraction models utilize the grammatical rules and the semantic context of language, while our model takes advantage of the identical semantic features determined by the aircraft movement pattern.
    \begin{figure}[H]
    	\centerline{\includegraphics[width=0.9\linewidth]{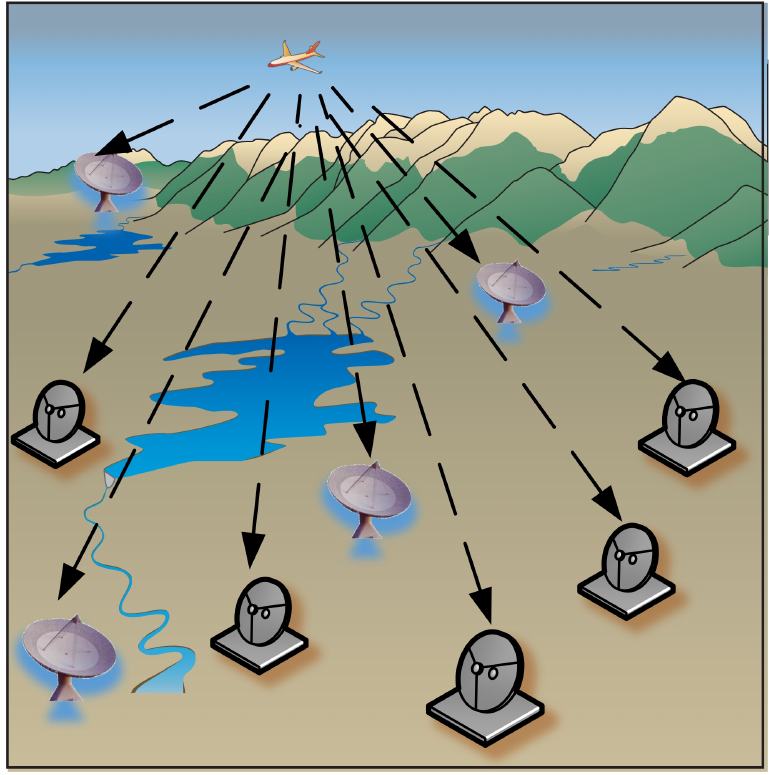}}
    	\caption{Heterogeneous radar network}
    	\vspace{-0.3cm}
    	\label{systemmodel}
    \end{figure}
    \par

    \subsection{System and Signal Models}
    As shown in Fig. \ref{systemmodel}, a heterogeneous radar network consists of $N$ spatially distributed radars, generally differing in carrier frequency, bandwidth, and pulse duration. When an aircraft passes, radars detect them independently and then output heterogeneous RCS signals which contain diverse grammatical features of the aircraft in the form of time-varying patterns.

    \par
    It is well known that the time-varying pattern of the RCS is determined by the characteristics of both the aircraft and radar, i.e., the movement pattern and electromagnetic shape of the aircraft, the carrier frequency, bandwidth, scan frequency, and pulse duration of the radar radio wave, and the position relationship between the aircraft and radar \cite{zhang2019rcs}.
    \begin{figure}[H]
    	\centerline{\includegraphics[width=\linewidth]{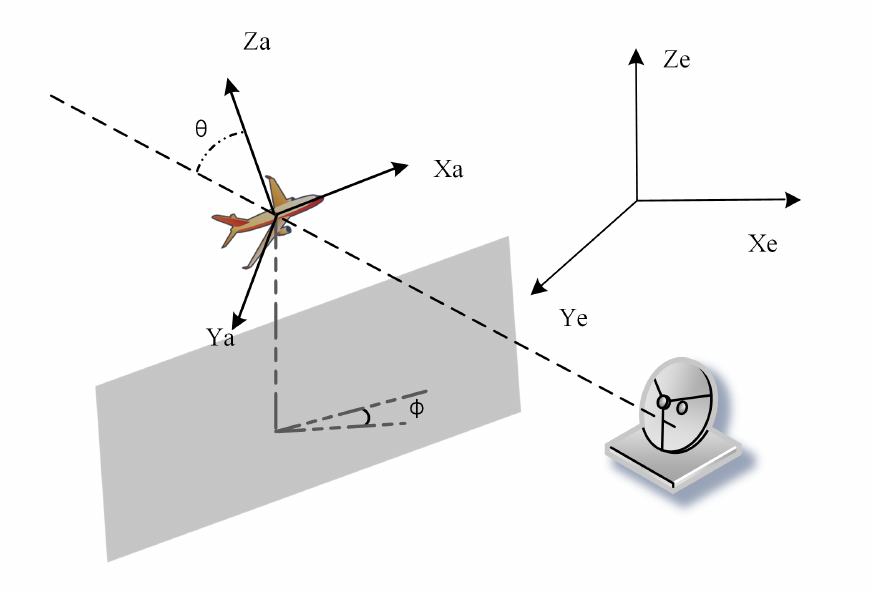}}
    	\caption{Model of aircraft and radar sight angle.}
    	\vspace{-0.3cm}
    	\label{Model1}
    \end{figure}
    \section{PRELIMINARY}
    \par
    To simplify the modeling of the RCS signals, two coordinates are used to describe the mass trajectory and micro-motion of the aircraft, i.e., the aircraft-body coordinate frame $({X_a},{Y_a},{Z_a})$ for mass trajectory from the viewpoint of the radar and the radar coordinate frame  $({X_e},{Y_e},{Z_e})$ for micro-motion, as shown in Fig. \ref{Model1}. The received RCS signal at time $t$ by the $n$-th radar can be represented by \begin{equation}\label{1}
    {m_n}{\rm{(}}t,f{\rm{)=}}{m_n}{\rm{(}}{\theta_n}{\rm{(}}t,f{\rm{),}}{\phi _n}{\rm{(}}t,f{\rm{))}},\
    \end{equation}
    where $f$ is the carrier frequency of the radar, ${\theta _n}$ and ${\phi _n}$ are, respectively, the horizontal and vertical components of the radar line-of-sight angle, which are defined by
    \begin{equation}\label{2}
    {\phi _n}(t) =\rho\cdot\arccos\frac{{{\Delta{y_n}}(t)}}{{\Delta{x_n}(t)}}   + \sqrt {1 - \rho }\cdot \sigma_\phi \cdot {\rm norm}(t + 1),\
    \end{equation}
    \begin{equation}\label{3}
    	\begin{aligned}
    		{\theta _n}(t) = &\rho\cdot\arccos \frac{{\Delta{z_n}(t)}}{{\sqrt {\Delta{x_n^2(t)} + \Delta{y_n^2(t)} + \Delta{z_n^2(t)}} }}  \\
    		&+ {\sigma _\theta } \sqrt {1 - \rho }\cdot {\rm norm}(t + 1),\
    	\end{aligned}
    \end{equation}
    where, ${\rm norm}(t + 1))$ denotes the normal distribution over $[-1,1]$, $\rho$ is the correlation coefficient, ${\sigma _\phi }$ and $\sigma _\theta$ are the disturbance variances of the horizontal and vertical angles, respectively.  $\Delta{x_n(t)}, \Delta{y_n(t)}$ and $\Delta{z_n(t)}$, the relative distances between the $n$-th radar and the aircraft in the aircraft-body coordinate frame, are determined by the mass trajectory and can be given by
    \begin{equation}\label{4}
    \begin{aligned}
    	&\left[ {\begin{array}{*{20}{c}}
    			\Delta{x(t)}\\
    			\Delta{y(t)}\\
    			\Delta{z(t)}
    	\end{array}} \right] = \left[ {\begin{array}{*{20}{c}}
    			1&0&0\\
    			0&{\cos \eta }&{\sin \eta }\\
    			0&{ - \sin \eta }&{\cos \eta }
    	\end{array}} \right] \cdot \left[ {\begin{array}{*{20}{c}}
    			{\cos \gamma }&0&{ - \sin \gamma }\\
    			0&1&0\\
    			{\sin \gamma }&0&{\cos \gamma }
    	\end{array}} \right]   \\
    	&\cdot \left[ {\begin{array}{*{20}{c}}
    			{\cos \varpi }&{\sin \varpi }&0\\
    			{ - \sin \varpi }&{\cos \varpi }&0\\
    			0&0&1
    	\end{array}} \right] \cdot \left[ {\begin{array}{*{20}{c}}
    			{{x_n}}&{ - {x_a}(t)}\\
    			{{y_n}}&{ - {y_a}(t)}\\
    			{{z_n}}&{ - {z_a}(t)}
    	\end{array}} \right],
    \end{aligned}
    \end{equation}
    where $\eta $, $\gamma $, $\varpi $ represent the yaw angle, pitch angle, and roll angle of the aircraft. $({x_n}, {y_n}, {z_n})$ and $({x_a}(t),{y_a}(t),{z_a}(t))$ represent the coordinates of the $n$-th radar and the aircraft in the radar coordinate frame, respectively.

    \par
    A cone-shaped aircraft \cite{2011Pseudo} usually spins to maintain flight stability, which is one of the most important characteristic micro-motions affecting the aspect angle.
    \par
    In another word, the time-varying pattern of RCS varies with the change of spatial observation position and the radar carrier frequency.
    \begin{figure}[H]
    	\centerline{\includegraphics[width=\linewidth]{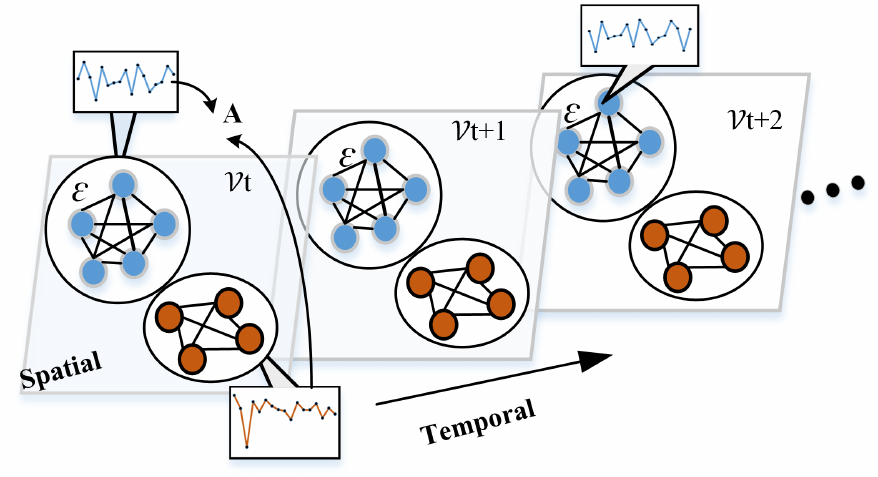}}
    	\caption{Illustration of RCS varying with time and space, where the radars with the same color constitute a homogeneous radar sub-net.}
    	\vspace{-0.3cm}
    	\label{graph}
    \end{figure}
    \par
    Aircraft recognition aims to classify target aircraft from the RCS signals received by the heterogeneous radar network. As shown in Fig. \ref{graph}, the heterogeneous radar network with $N$ radars can be viewed as an $N$-node undirected graph with weights $\mathcal{G} = (\mathcal{V}, \mathcal{E}, \bm{A})$, where $\mathcal{V}$ consists of RCS vectors with $\left| {\mathcal{V}} \right|{\rm{ = N}}$. $\mathcal{E}$ is an edge set where an edge represents the connected nodes within the same sub-net, and $\bm{A} \in {\mathbb{R}^{N \times N}}$ is the weight adjacency matrix whose elements represent the similarities between the RCS signals. We train a GNN model on $\mathcal{G}$ to learn a nonlinear complex function $g( \cdot )$, which can map graph signal vectors to labels such that the target aircraft can be correctly classified.

    \begin{figure}[H]
    	\centerline{\includegraphics[width=0.95\linewidth]{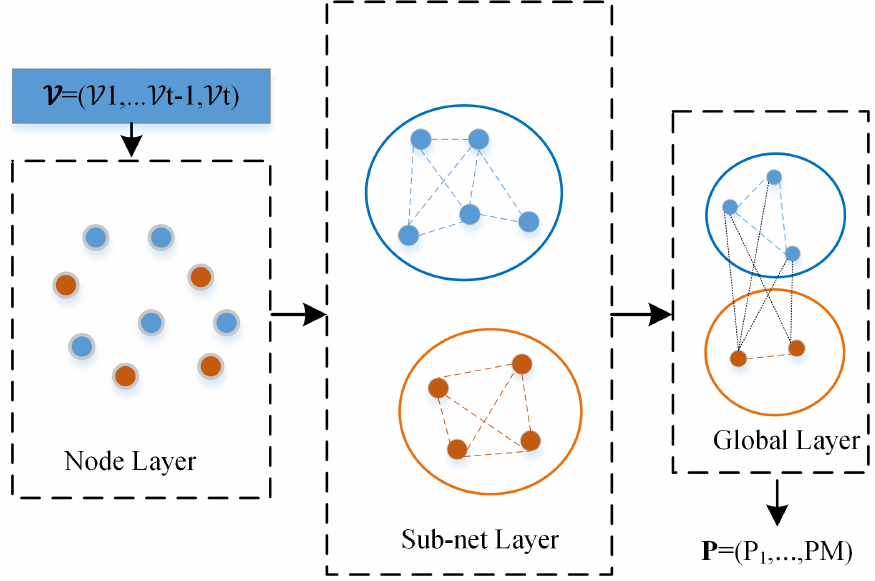}}
    	\caption{Illustration of the STFGACN model, which consists of a node layer, a subnet layer, and a global layer.}
    	\vspace{-0.3cm}
    	\label{ModelI}
    \end{figure}

    \section{PROPOSED MODEL}
\begin{figure*}[t]
    	\centerline{\includegraphics[width=\linewidth]{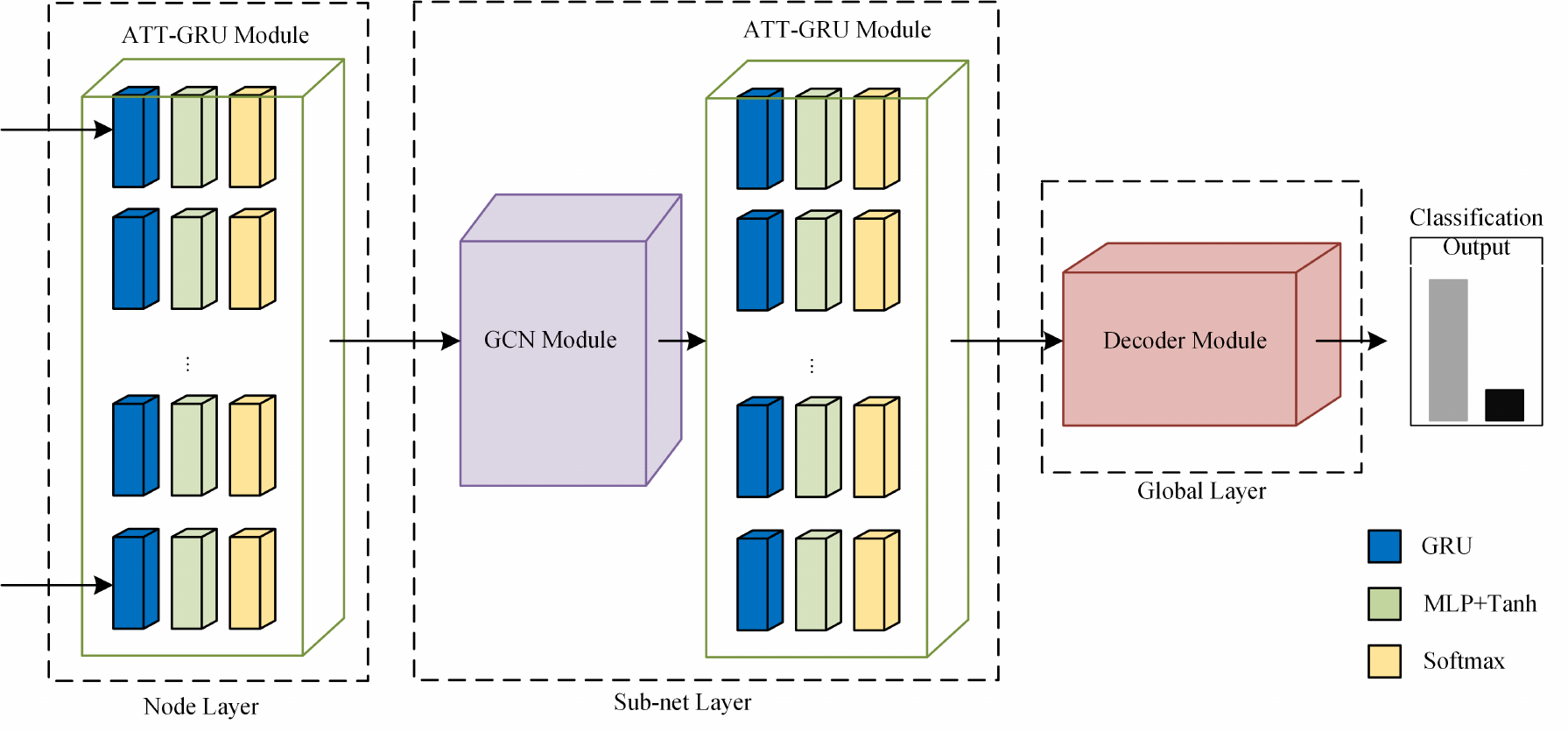}}
    	\caption{Structure of the spatio-temporal-frequency graph attention convolutional network (STFGACN). }
    	\vspace{-0.3cm}
    	\label{structure}
    \end{figure*}

    In this section, we elaborate on the proposed STFGACN. As shown in Fig. \ref{ModelI}, the structure of the proposed model consists of three layers, i.e., a node layer, a subnet layer, and a global layer. In the node layer, each node independently extracts temporal features as a sample in the spatial-frequency domain. In the node layer, nodes with the same frequency constitute a subnet, and the spatial-temporal features are extracted and fused as a sample in the frequency domain. In the global layer, subnets are converged and semantic features are distilled from the spatial-temporal-frequency domain. The process in each layer is detailed one by one, and the overall network structure of the proposed model is illustrated in Fig. \ref{structure}. In the node layer, gate recurrent unit (GRU) with attention mechanism referred to as ATT-GRU is employed to extract temporal features at each node. In the sub-net layer, the graph convolution network (GCN) and the ATT-GRU module are employed to extract temporal-spatial features from each homogeneous radar subnet. In the global layer, a decoder module is designed to distill semantic features from the temporal-spatial-frequency domain for accurate aircraft recognition.

    \begin{figure*}[t]
    	\centerline{\includegraphics[width=\linewidth]{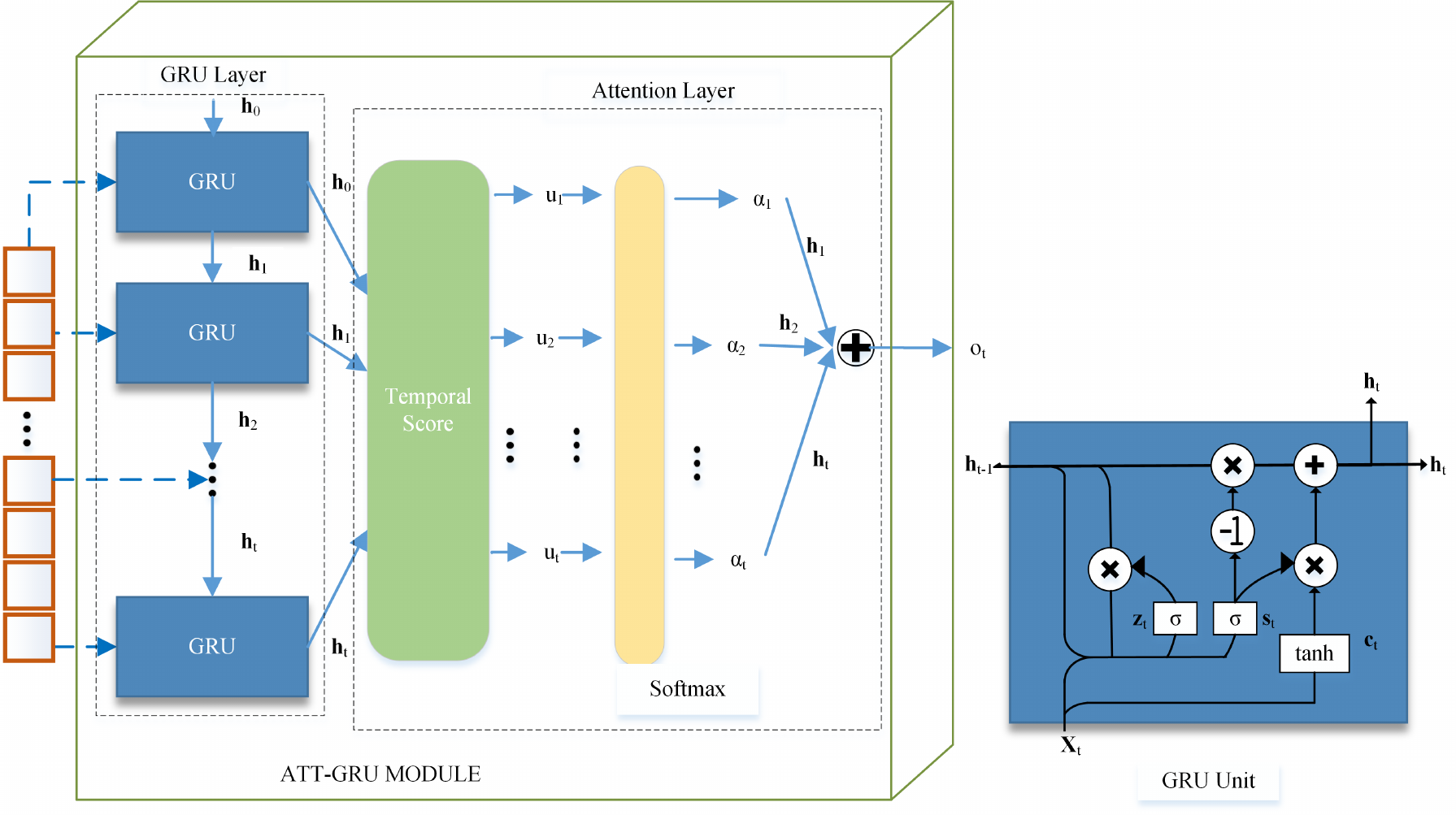}}
    	\caption{Schematics of the layer. (a) Structure of the ATT-GRU module. (b) Structure of the GRU unit.}
    	\vspace{-0.3cm}
    	\label{Model2}
    \end{figure*}

    \subsection{Node Layer}
    As illustrated above, the time-varying pattern of RCS is determined by both the observed movement pattern of aircraft and the ranging pattern of radar, which are related to semantic and grammar features, respectively. The features in the RCS are the mass trajectory and micro-motion features. The micro-motion characteristics are related to high-frequency features, while the mass trajectory characteristics are related to low-frequency features. In order to extract both long-term and short-term features from RCS, the GRU with the attention mechanism termed ATT-GRU is employed, whose structure is illustrated in Fig. \ref{Model2}.
    \par
    In the proposed ATT-GRU module, the GRU unit \cite{cho2014learning} consists of update gates $\bm{s}_t$ and reset gates $\bm{z}_t$ for remembering or forgetting historic information. These two gates update the hidden state $\bm{h}_t$ through a linear combination of the candidate state $\bm{c}_t$ and previous hidden state  $\bm{h}_{t-1}$ as follows
    \par
    \begin{equation}\label{5}
    	\bm{s}_t = \sigma (\bm{W}_s\bm{x_}t + \bm{V}_s\bm{h}_{t-1} + \bm{b}_s)\,
    \end{equation}
    \begin{equation}\label{6}
    	\bm{z}_t = \sigma (\bm{W}_z\bm{x}_t + \bm{V}_z\bm{h}_{t-1} + \bm{b}_z)\,
    \end{equation}
    \begin{equation}\label{7}
    	\bm{c} _t = \tanh (\bm{W}_c\bm{x}_t + \bm{V}_c(\bm{h}_{t-1} \otimes \bm{z}_t+ \bm{b}_c))\,
    \end{equation}
    \begin{equation}\label{8}
    	\bm{h}_t= \bm{s}_t \otimes \bm{c} _t + \bm{h}_{t-1} \otimes (1 - {\bm{s}_t})\,
    \end{equation}
    where $\sigma ( \cdot )$ denotes the sigmoid function, $\bm{W}_s$, $\bm{V}_s$, $\bm{W}_z$, $\bm{V}_z$, $\bm{W}_c$, $\bm{V}_c$ are the weight matrices, and $\bm{b}_s$, $\bm{b}_z$, $\bm{b}_c$ are the bias vectors.
    \par
    In order to extract long-term features, the attention mechanism is incorporated, which weighs the temporal features with different periods adaptively. The weight at the $i$-th time step is calculated from hidden state $\bm{h}_i$ by
    \begin{equation}\label{9}
    	{u_i}{\rm{ = }}\tanh (\bm{W}_u\bm{h}_i + \bm{b}_u),\
    \end{equation}
    where $\bm{W}_u$ and $\bm{b}_u$ are the weight matrices and bias, respectively. Weight ${\alpha _i}$ is obtained by normalizing the attention weights $u_i$ by the softmax function as follows
    \begin{equation}\label{10}
    	{\alpha _i}{\rm{ =  softmax(}}{u_i}{\rm{)  =  }}\frac{{\exp ({u_i})}}{{\sum\nolimits_{j = 1}^K {\exp ({u_j})} }}.\
    \end{equation}

    \par
    Then $\bm{h}_i$ is weighted by ${\alpha _i}$ to yield the temporal features ${o_t}$ as follows
    \begin{equation}\label{11}
    {o_t} = \sum\limits_{i = 1}^K {{\alpha _i}\bm{h}_i} .\
    \end{equation}

    \begin{figure}[H]
    	\centerline{\includegraphics[width=\linewidth]{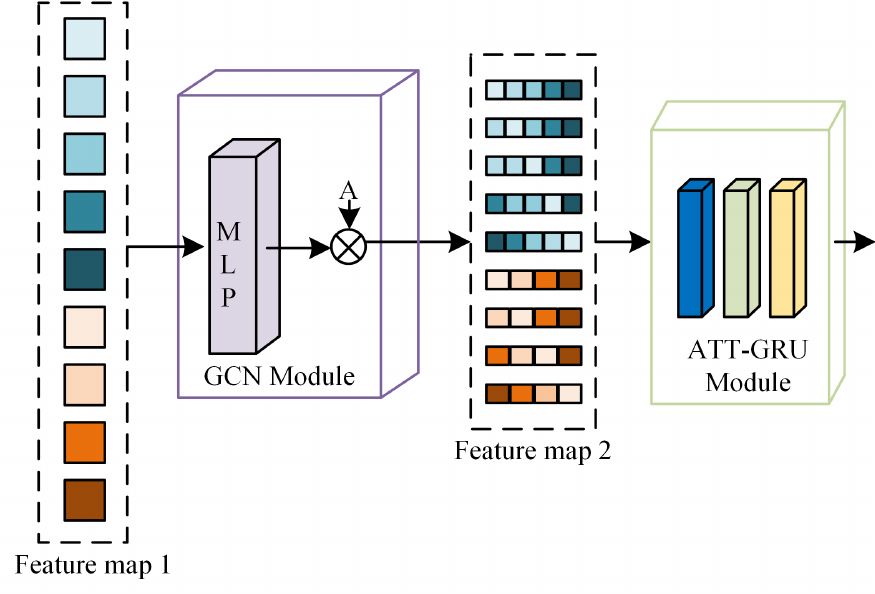}}
    	\caption{Illustration of the Sub-net layer.}
    	\vspace{-0.3cm}
    	\label{Sub-Net}
    \end{figure}
    \subsection{Sub-net Layer}
    In the proposed model, a subnet is composed of nodes with the same carrier frequency, on which spatial-temporal convolution is applied to extract spatial-temporal features as sampled features in the frequency domain. To avoid the computation-intensive Fourier transform in the graph convolution, 1stChebNet GCN \cite{kipf2016semi} is adopted so that we can build a deeper network by replacing the explicit parameterization via, e.g., the Chebyshev polynomials \cite{defferrard2016convolutional} with computation-reduced first-order polynomial, which is expressed by
    \begin{equation}\label{12}
    {g }{ _G}o \approx \bm{W}_G {\rm{(}}{{\bm{I}}_{n}}{\rm{ + }}{{\bm{D}}^{{\rm{ - }}\frac{{\rm{1}}}{{\rm{2}}}}}{\bm{A}}{{\bm{D}}^{{\rm{ - }}\frac{{\rm{1}}}{{\rm{2}}}}}){\bm{O}},
    \end{equation}
    where ${g }{ _G}$ denotes the graph convolution operation based on the spectral graph convolution,  ${\bm{D}} \in {\mathbb{R}^{N \times N}}$ is a diagonal degree matrix, $\bm{W}_G$ is the weighting coefficient, ${\bm{A}}$ is the weight adjacency matrix, and ${{\bm{I}}_{n}}$ is an identity matrix.
    \par
    The simplified graph convolution operator can act as a temporal-spatial filter to distill local information by only taking neighboring nodes into consideration. By stacking the graph convolution filters, information from multi-step neighboring nodes is converged, which achieves the same effect as K-localized convolution but with a significantly reduced computational complexity.
    \par
    Fig. \ref{Sub-Net} illustrates the proposed GCN module, which fuses spatial semantic features via the following tensor multiplication
    \begin{equation}\label{13}
    	{g }{ _G}o {\rm{ = }} \bm{W}_G {\widehat{\bm{A}}}{\bm{O}},
    \end{equation}
    where $\widehat{\bm{A}}{\rm{ = }}{{\bm{I}}_{\rm{n}}}{\rm{ + }}{{\bm{D}}^{{\rm{ - }}\frac{{\rm{1}}}{{\rm{2}}}}}{\bm{A}}{{\bm{D}}^{{\rm{ - }}\frac{{\rm{1}}}{{\rm{2}}}}}$ is the normalized adjacency matrix. Every node is updated by aggregating information of its neighborhood nodes. The fused spatial-temporal information is further distilled by the following ATT-GRU module which shares the same structure as the one at node layer.

    \subsection{Global Layer}
\begin{figure}[H]
    	\centerline{\includegraphics[width=\linewidth]{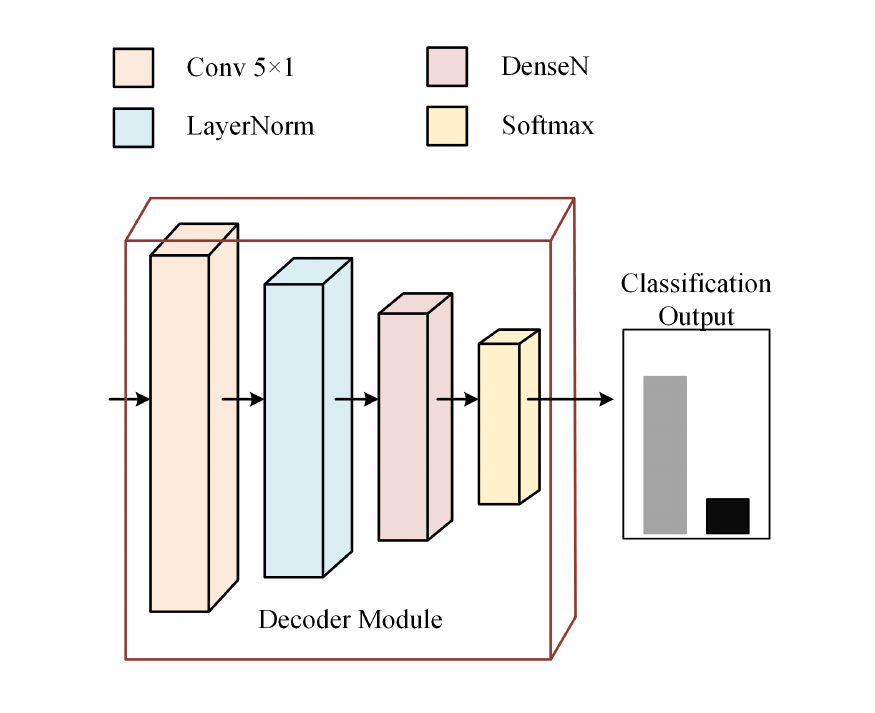}}
    	\caption{Decoder structure in the global layer.}
    	\vspace{-0.3cm}
    	\label{Decoder}
    \end{figure}

    In order to better extract semantic features determined by the movement pattern, a decoder module is designed to converge information from subnets, and then fuse temporal-spatial-frequency features by a CNN module, whose structure is depicted in Fig. \ref{Decoder}. The decoder begins with a 5$\times$1 convolution operation to extract semantic features from the feature-map yielded by the sub-net layer. Then, layer normalization is peformed to prevent overfitting before higher semantic feature extraction by a fully connected dense layer. The softmax activation function is used to generate aircraft classification results.

    \begin{figure*}
    	\centering
    	\subfloat[ ]{
    		\includegraphics[width=0.45\linewidth]{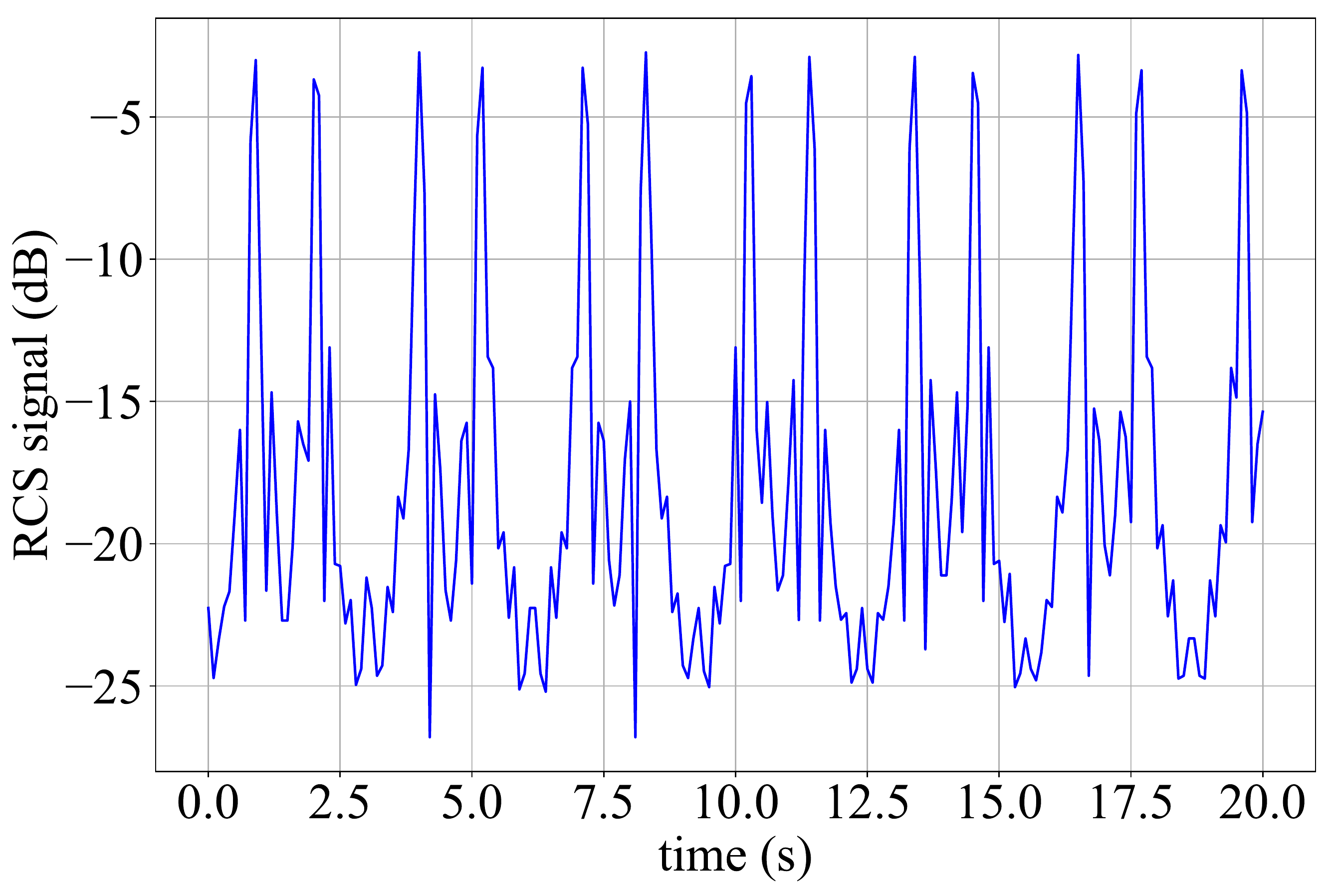}}
    	\hspace{0.01\linewidth}
    	\subfloat[ ]{
    		\includegraphics[width=0.45\linewidth]{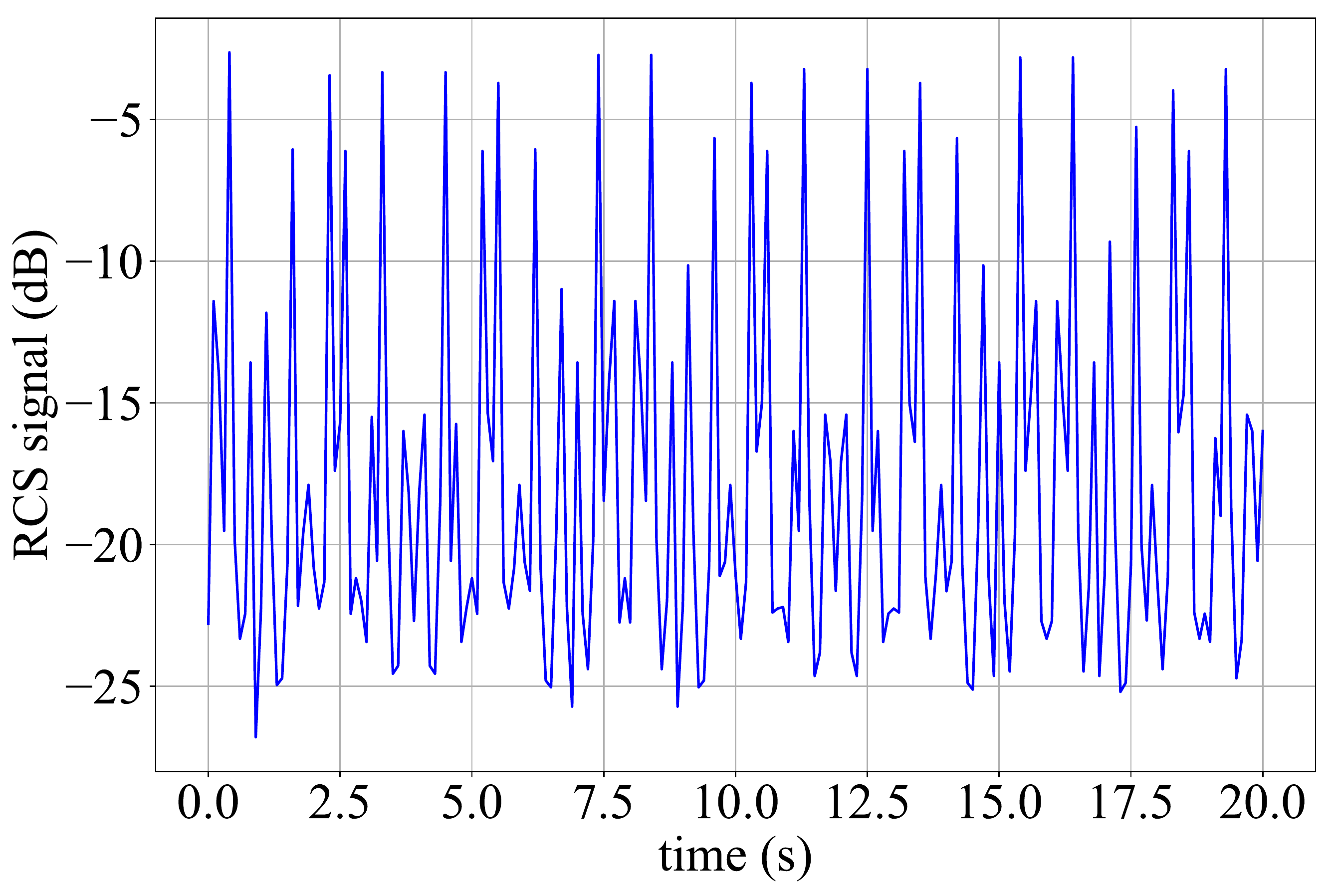}}
    	\vfill
    	\subfloat[ ]{
    		\includegraphics[width=0.45\linewidth]{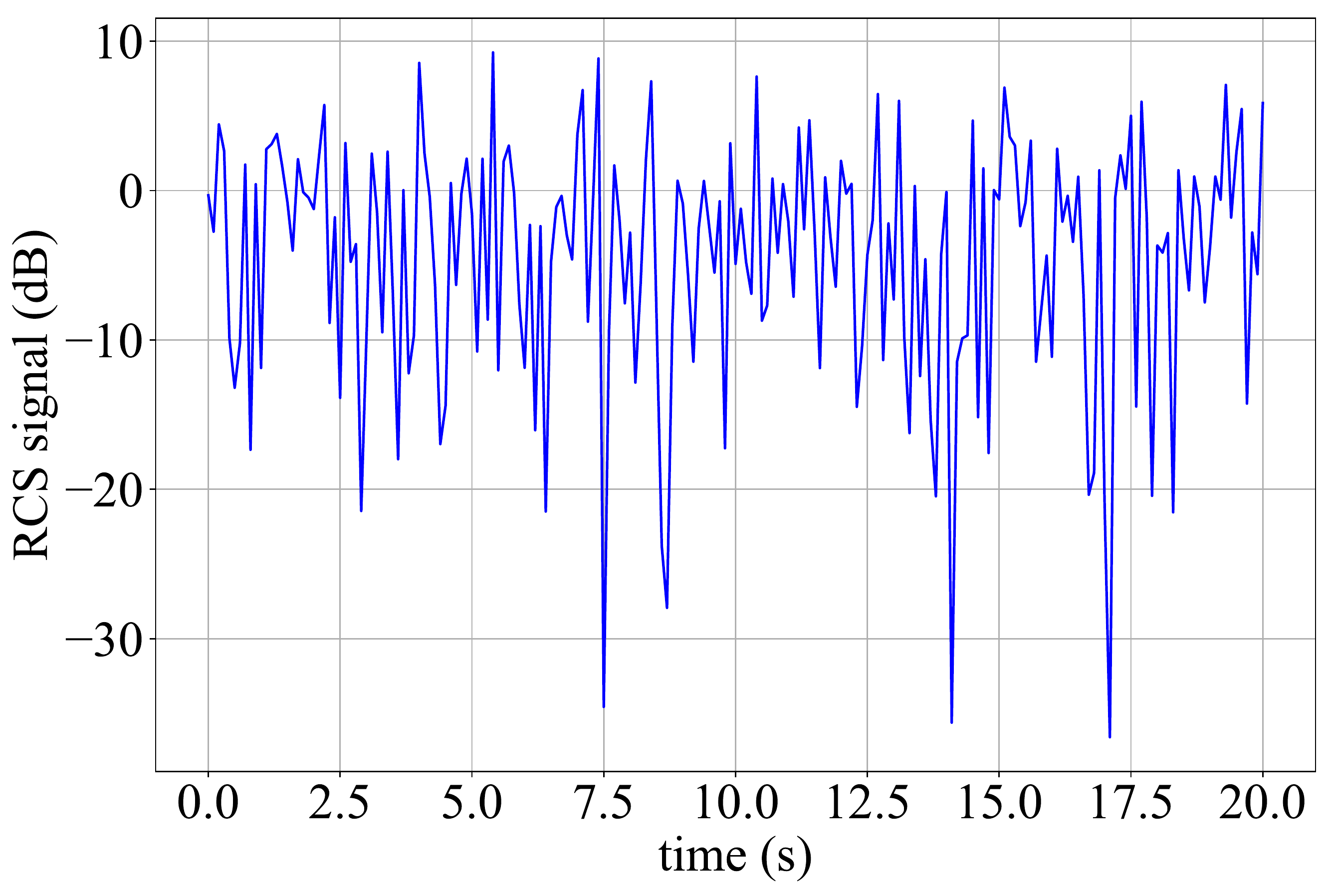}}
    	\hspace{0.01\linewidth}
    	\subfloat[ ]{
    		\includegraphics[width=0.45\linewidth]{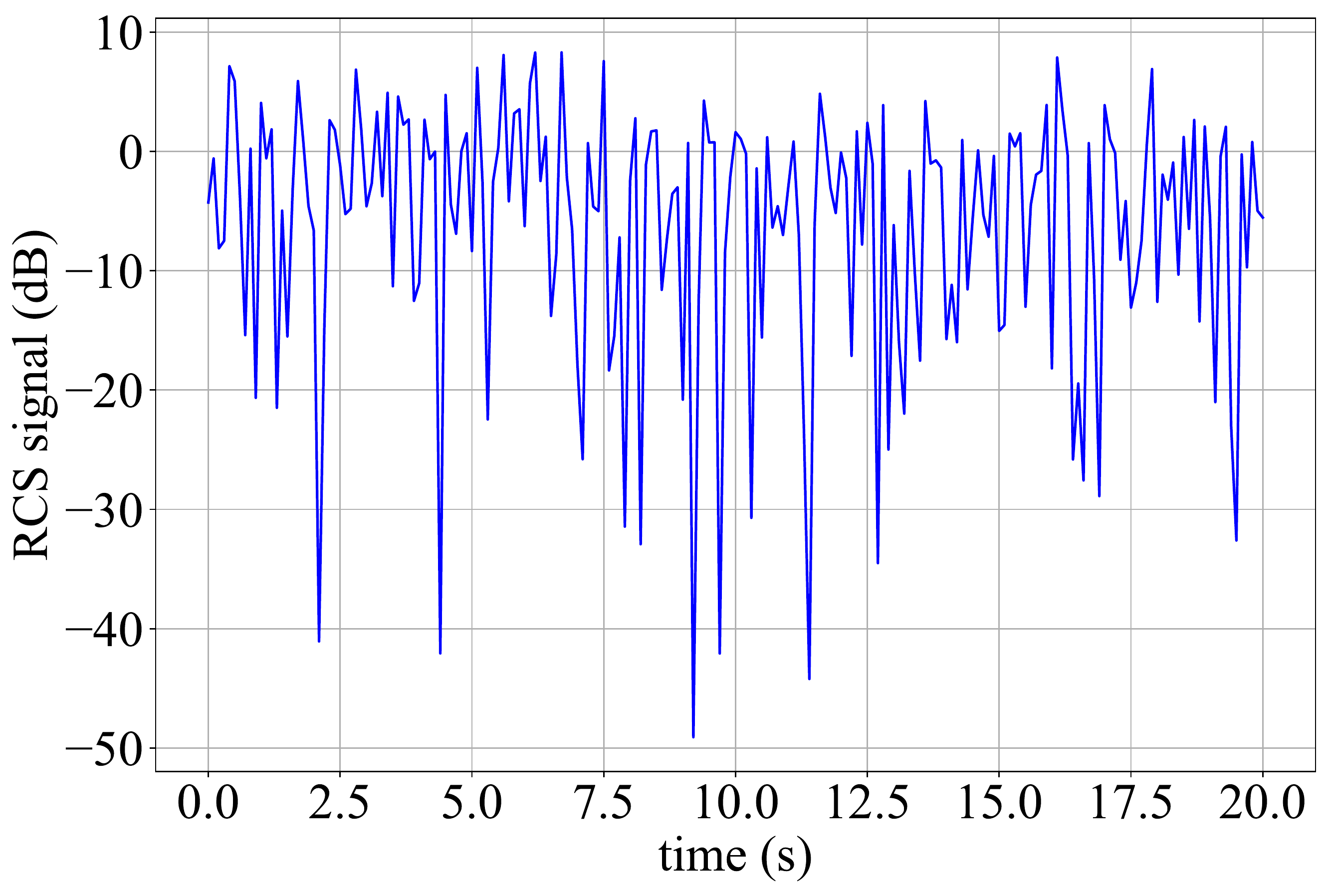}}
    	\caption{RCS signals.  (a) type-1 aircraft without noise; (b) type-2 aircraft without noise; (c) type-1 aircraft at the SNR of 10\,dB; and (d) type-2 aircraft at the SNR of 10\,dB.}
    	\label{RCSSignal}
    \end{figure*}

    \section{NUMERICAL EXPERIMENTS}

    \subsection{Experiment System}
    In our experiments, $N$ = 9 spatially distributed radars are divided into two subnets, each of which has different carrier frequencies. Specifically, the first subnet consists of 5 radars opearting at 6.25\,GHz, while the other four radars operate at 1.52\,GHz to form the second subnet. Two types of aircrafts are considered in our experiments, which have the same EM scatter shape and the mass trajectory but different micro-motions. The parameter settings of the aircrafts are listed in Table \ref{table1}, and the channel model is the additional white Gaussian noise (AWGN) channel, which is widely adopted when the pitch angle of the aircraft is large.
    \par
    The RCS signals are obtained through EM simulation using the CST simulation software \cite{2015RCS}, and the simulated RCS signals are illustrated in Fig. \ref{RCSSignal}, from which we can easily distinguish these two types of aircrafts from the RSC without the AWGN. However, even when the noise is moderate, we cannot classify the aircrafts from the RCS signals directly.
    \begin{table}[H]
    	\centering
    	\caption{The parameters of the aircraft.}
    	\begin{tabular}{c|p{1.565em}|p{6.5em}|p{6.5em}}
    		\hline
    		\multicolumn{2}{r|}{} & \textbf{type-1 aircraft} & \textbf{type-2 aircraft} \bigstrut\\
    		\hline
    		\multicolumn{2}{p{7.44em}|}{\textbf{Mass trajectory}} & \multicolumn{2}{p{10.315em}}{5km/s} \bigstrut\\
    		\hline
    		\multicolumn{1}{c|}{\multirow{2}[4]{*}{\textbf{micro-motion}}} & $f_1$    & 0.64\,Hz & 2.75\,Hz \bigstrut\\
    		\cline{3-4}          & $f_2$    & 1.67\,Hz & 8.72\,Hz \bigstrut\\
    		\hline
    	\end{tabular}%
    	\label{table1}%
    \end{table}%
    \subsection{Data Processing}
    1000 RCS signal segments are simulated from each radar for each aircraft, and then 18,000 RCS signal segments are obtained as the sample set. An RCS signal segment lasts 10 seconds with 200 samples. The dataset is divided into the train, test and validation sets with a ratio of 7:2:1.
    \par
    The data is normalized by the maximum and minimum normalization method. According to the spatial distribution of the radars, the adjacency matrix ${\bm{A}}$ can be constructed, whose $(i,j)$-th element $a_{ij}$ is given by
    \begin{equation}\label{50}
    	{{\rm{a}}_{ij}} = \left\{ {\begin{array}{*{20}{c}}
    			{0.3 + \frac{1}{{{d_{ij}}}},f{c_i} = f{c_j}}\\
    			{\frac{1}{{{d_{ij}}}},otherwise},
    	\end{array}} \right.
    \end{equation}
    where $f{c_i}$ and $f{c_j}$ denote the operating frequencies of node $i$ and node $j$, respectively. $d_{ij}$ indicates the distance between these two radars.

    \subsection{Model Parameter Setting and Environment}
    The experiments are implemented on a Linux server with CPU: Intel(R) Xeon(R) CPU E5-2630 v3 @ 2.40GHz, and GPU: NVIDIA GeForce GTX 1080.

    \begin{itemize}
    	\item \textbf{Performance Metrics:}
    	\par
    	Four widely used performance metrics are employed, i.e., Accuracy, Precision, Recall and F1-Score, which are defined as
    	\begin{equation}\label{51}
    		Accuracy = \frac{{TP + TN}}{{TP + TN + FP + FN}},\
    	\end{equation}
    	
    	\begin{equation}\label{52}
    		Precision = \frac{{TP}}{{TP + FP}},\
    	\end{equation}
    	
    	\begin{equation}\label{53}
    		Recall = \frac{{TP}}{{TP + FN}},\
    	\end{equation}
    	
    	\begin{equation}\label{54}
    		F1 = 2\frac{{Precision \times Recall}}{{Precision + Recall}},\
    	\end{equation}
    	where $TP$ is the number of correctly recognized positive samples (the first aircraft type), $TN$ is the number of correctly recognized negative samples (the second aircraft type), $FP$ is the number of incorrectly recognized positive samples, and $FN$ is the number of incorrectly recognized negative samples.
    	
    	\item \textbf{Hyperparameters:}
    	\par
    	In temporal attention mechanisms, the input unit and hidden unit of the GRU are set to 200 and 64, respectively. The size of the graph convolution kernel is set to 1 in the model with the first-order approximation. We train our model by the Adam optimizer [30] with an initial learning rate of 0.001. The Batch size is 5 and the step decay rate is 0.5. The loss function is SigmoidCrossEntropyLoss. The training process stops after 100 epochs or 10 non-improving validation loss epochs.
    	\item \textbf{Baseline:}
    	\par
    	The STFGACN is compared with four baseline methods: (1) FFT \cite{swami2000hierarchical}; (2) SVM \cite{gokkaya2019novel}; (3) GRU \cite{cho2014learning}; and (4) STGCN \cite{yu2017spatio}.
    	\par
    	These reference methods can be categorized into two groups. Group 1 includes STGCN and STFGACN, which are based on the radar network, while group 2 consists of the other three models, which are single radar-based detection methods. For a fair comparison, the decision fusion strategy is employed for the group 2 methods. That is, a voting classifier is constructed where the nine single radar-based models act as element classifiers. These ensemble classifiers are denoted by FFT(9), SVM(9), and GRU(9) in the following.
    \end{itemize}
    \begin{figure}[H]
    	\centerline{\includegraphics[width=\linewidth]{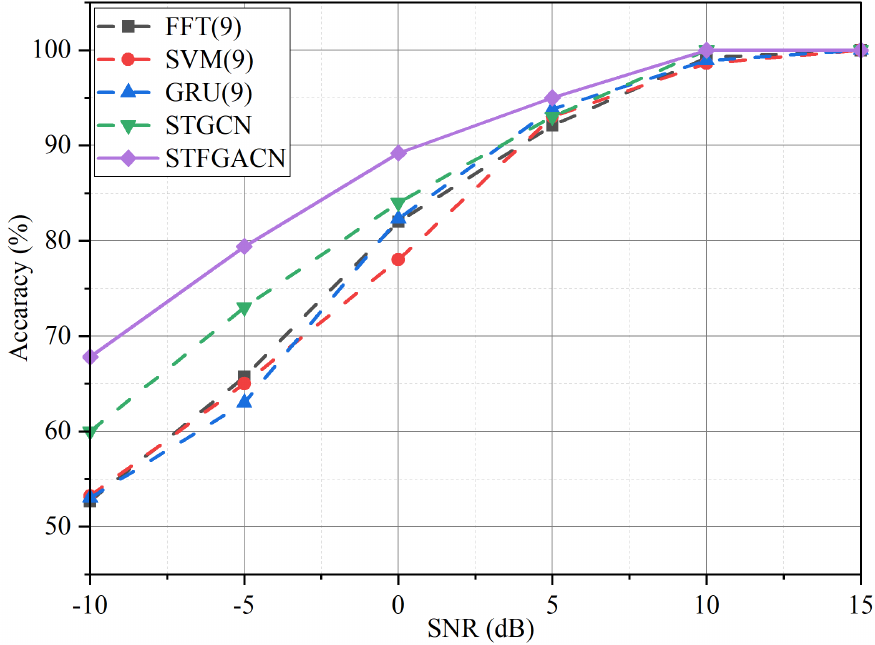}}
    	\caption{Accuracy of aircraft recognition versus the SNR for all concerned methods.}
    	\vspace{-0.3cm}
    	\label{Effect}
    \end{figure}	
    \begin{figure}[H]
    	\centerline{\includegraphics[width=\linewidth]{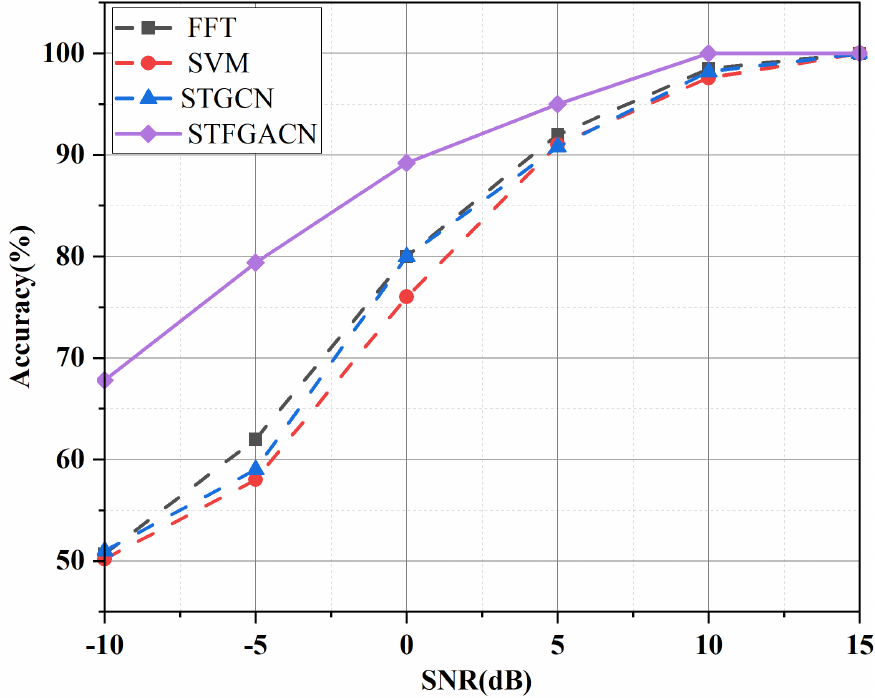}}
    	\caption{Accuracy performance comparison between the GCN-type methods and the single radar-based methods without voting policy.}
    	\vspace{-0.3cm}
    	\label{Effect1}
    \end{figure}
    \subsection{Experimental Results}
    Fig. \ref{Effect} shows the experimental results of all concerned model. The following observations can be made from Fig. \ref{Effect}:
    \par
    1) The proposed STFGACN outperforms all reference models, and the performance gain increases with the decrease of the SNR, which verifies that the proposed STFGACN method can greatly enhance the reliability of aircraft recognition, especially in the low SNR region where traditional methods degrade severely. The performance gain of the proposed method stems from the spatial-temporal-frequency domain feature extraction and semantic feature fusion, which is testified by the following experiments.
    \par
    2) The GNN-based models are able to achieve better aircraft recognition than single radar-based counterparts, especially in the low SNR region. This result shows that the GNN-based methods can extract and fuse more useful information from multiple domains. Although the voting policy is deployed for the utilization of multi-domain information, the group 2 methods have much lower efficiency in terms of fully exploiting the multi-domain information, which is further validated by the following experiments when comparing the proposed method with the single radar-based methods without voting policy. The experimental results are shown in Fig. \ref{Effect1}.
    \par
    3) The results in Fig. \ref{Effect1} show that the performance enhancement is subtle when fusing the detection results from nine radars and then yielding the final output via the voting policy for the single radar-based methods. Specifically, at SNR = -10\,dB, the accuracy increases from 50.7\%, 50.3\%, 51.1\% to 51.6\%, 53.2\%, 53\% for FFT, SVM, and GRU, respectively. In other words, the decision-level fusion scheme for the single radar-based methods achieve no more than 3\% performance gain, which indicates that the decision-level information fusion cannot achieve effective semantic feature distillation.
    \par
    4) For the GNN-based methods, the proposed STFGACN can achieve about a 2\,dB SNR gain over the STGCN model in the low SNR region. This SNR gain is attributed mainly to the temporal feature extraction by the ATT-GRU module, where a CNN is employed in the STGCN. As we know, ATT-GRU can better distill large-spanned temporal patterns than CNN, which is good at local feature distillation.
    \par
    5) At higher SNR scenarios, all methods can reliably recognize aircrafts. That is, the accuracy of aircraft recognition is above 90\% when the SNR is larger than 5\,dB, and the accuracy approaches 100\% when the SNR is above 10\,dB. However, in practical applications, low SNRs are more representative scenarios. Given the threshold of recognition accuracy, an SNR gain means an increased radar detection range. For example, given the accuracy threshold of 75\%, the proposed method can achieve a 4.7\,dB SNR gain over the single radar-based methods, which is equivalent to a three-fold increase of the detection range.

    \subsection{Ablation Experiments}

    % Table generated by Excel2LaTeX from sheet 'Sheet1'
    \begin{table}[H]
    	\centering
    	\caption{Comparison with variants of STFGACN.}
    	\begin{tabular}{c|cccc}
    		\hline
    		\textbf{Method} & \textbf{Accuracy} & \textbf{Precision} & \textbf{Recall} & \textbf{F1} \bigstrut\\
    		\hline
    		\textbf{GRU} & 51.00\% & 50.98\% & 51.80\% & 0.51 \bigstrut[t]\\
    		\textbf{ATT-GRU} & 53.00\% & 52.90\% & 54.80\% & 0.54 \\
    		\textbf{STGCN} & 61.00\% & 60.70\% & 62.40\% & 0.62 \\
    		\textbf{STFGACN-1F} & 63.20\% & 62.89\% & 64.40\% & 0.64 \\
    		\hline
    		\textbf{STFGACN-2F} & \textbf{67.80\%} & \textbf{67.78\%} & \textbf{68.60\%} & \textbf{0.68} \bigstrut[b]\\
    		\hline
    	\end{tabular}%
    	\label{table2}%
    \end{table}%
    To understand the contribution to detection accuracy by each function module of the proposed model, ablation experiments are conducted, which can clarify the contribution of information dimension from spatial-temporal-frequency domain.
    \par
    The compared variates of the proposed method include four models, i.e., ATT-GRU, STGCN, STFGACN-1F, STFGACN-2F.

    \begin{itemize}
    	\item \textbf{} The ATT-GRU model is the simplified version when only one radar is used and then it can only distill the temporal features of received signals. This model is also the enhanced version of GRU.
    	\item \textbf{} The STGCN model can extract temporal-spatial features by CNN module, which can be regarded as the reduced version of the proposed model in the temporal-spatial domain by replacing ATT-GRU with CNN.
    	\item \textbf{} The STGACN-1F model denotes the radar network consisting of the same frequency, and it compares with the reference STGCN model to verify the temporal-spatial feature extraction capability.
    	\item \textbf{} The STGACN-2F model is the one used for the multiple-frequency radar network.
    \end{itemize}
    The experimental results are listed in Table \ref{table2}, where all models work at SNR = -10\,dB in order to evaluate their robustness. We can observe from Table \ref{table2} : $i$) the attention mechanism can offer 2\% gain over the basic GRU model; $ii$) STFGACN-1F achieves 10.8\% accuracy gain compared to ATT-GRU, which testifies that the extra feature in the spatial domain can greatly improve the accuracy; $iii$) STFGACN-1F achieves 2.8\% accuracy gain over STGCN, which shows ATT-GRU can better extract features over large range; $iv$) STFGACN-2F can present 4.6\% accuracy gain than STFGACN-1F, which shows that extra information can be distilled from the frequency domain. It can be concluded from the above results that the spatial dimension information contributes to the accuracy improvement the most, and the semantic features from the temporal and spatial domains are also important for reliable recognition.
    \par
    In order to quantify the performance gain attributed to each constituent module, we present the performance of detection accuracy with varying SNR settings.

    \subsubsection{Information dimension expansion - Time dimension}
    \par
\begin{figure}[t]
    	\centerline{\includegraphics[width=\linewidth]{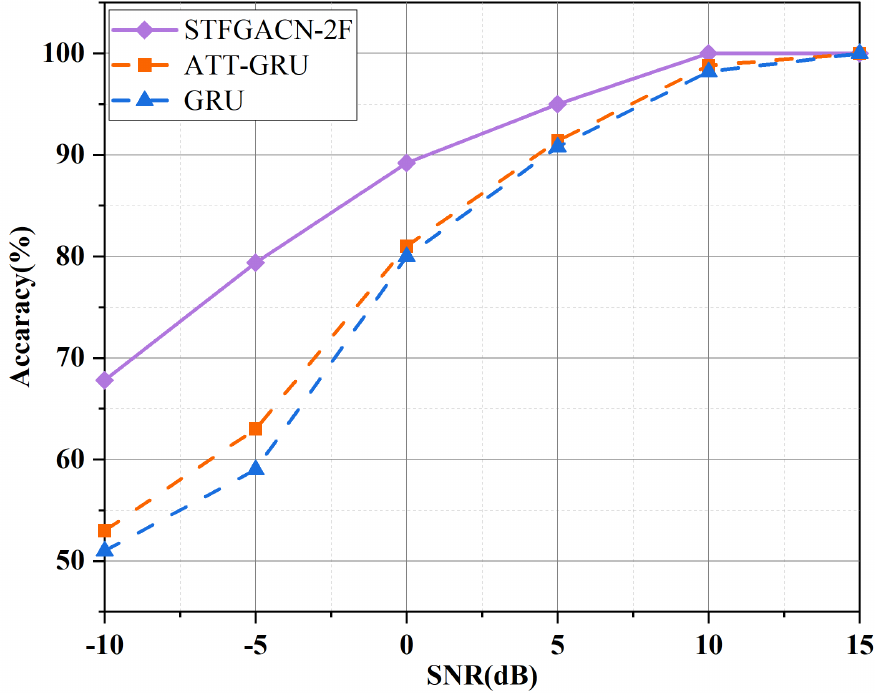}}
    	\caption{Temporal feature extraction comparison between ATT-GRU and GRU over a range of SNRs.}
    	\vspace{-0.3cm}
    	\label{Effect2}
    \end{figure}

    Fig. \ref{Effect2} plots the accuracy versus the SNR for the GRU, ATT-GRU and STFGACN-2F models. As can be seen from the figure, the attention mechanism can increase the accuracy by 3\% in the low SNR region. In the experiments, since the feature frequencies of micro-motion of the two types of aircrafts are 0.64\,Hz and 1.67\,Hz, respectively, the temporal periods are then about 1.56 seconds and 0.6 seconds, respectively, which span about 31 and 12 samples. The attention mechanism can extract more long-period features from the RCS. It can be seen that when the threshold of detection accuracy is 75\%, the ranging distance of the ATT-GRU is one and a half times of that of the GRU.

    \subsubsection{Information dimension expansion -  Spatial dimension}
    \par
\begin{figure}[t]
    	\centerline{\includegraphics[width=\linewidth]{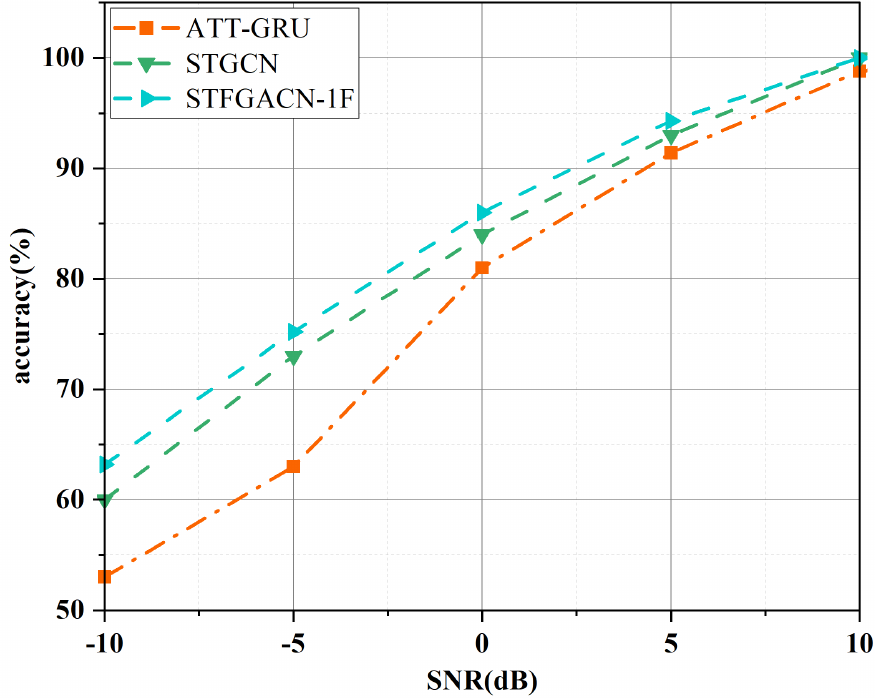}}
    	\caption{ATT-GRU versus STFGACN-1F in spatial feature extraction.}
    	\vspace{-0.3cm}
    	\label{Effect3}
    \end{figure}

    We present the experimental results of the ATT-GRU, STGCN and STFGACN-1F methods over a range of SNRs in Fig. \ref{Effect3} to show the performance gain attributable to extra spatial information. The results in Fig. \ref{Effect3} show that the extracted spatial information can greatly enhance the detection accuracy, especially in low SNR region. STFGACN-1F has an accuracy improvement of 10\% compared with ATT-GRU, when the SNR is below -5\,dB. It can be concluded that the RCS signals received by the spatially dispersed radars have strong spatial correlation. By extracting and then fusing spatial features, the proposed method can greatly enhance the detection performance in the low SNR region. When the detection accuracy threshold is 75\%, the ranging distance of our method can be doubled compared to its single radar-based counterpart.
    \begin{figure}[t]
    	\centerline{\includegraphics[width=\linewidth]{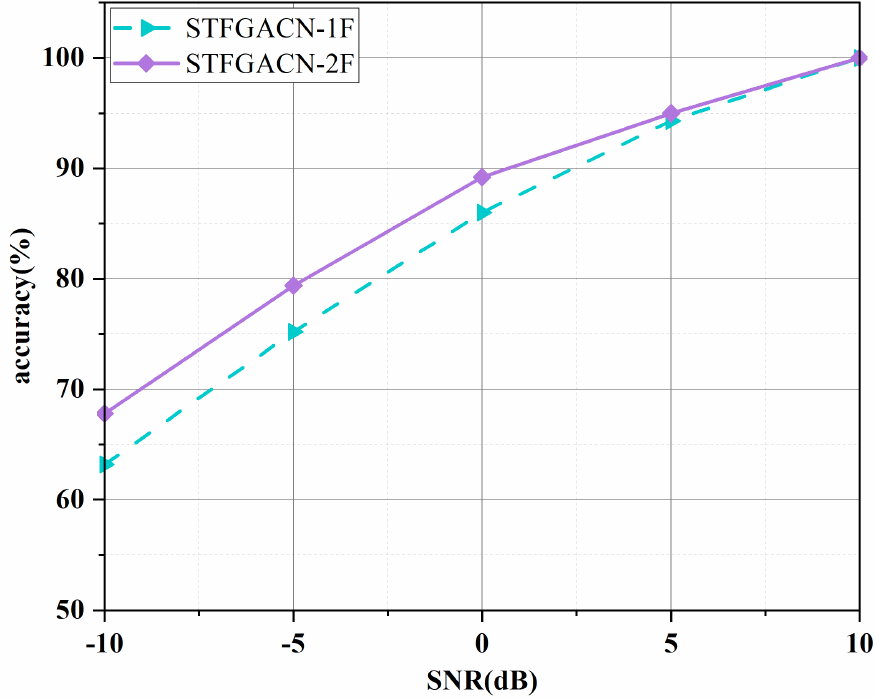}}
    	\caption{Performance comparison of the proposed STFGACN with the heterogeneous and homogeneous networks.}
    	\vspace{-0.3cm}
    	\label{Effect4}
    \end{figure}

    \subsubsection{Information dimension expansion - frequency dimension}
    \par
    The performance comparison of the proposed STFGACN with the heterogeneous and homogeneous networks is shown in Fig. \ref{Effect4}. By fusing features from multiple frequency domains, a 4.6\% accuracy improvement is achieved. In terms of the ranging distance, the proposed method improves about 60\% when the threshold of recognition accuracy is above 75\%.

    \section{Conclusion}
    In this paper, the Spatio-Temporal-Frequency Graph Attention Convolutional Network (STFGACN) model was proposed for aircraft recognition using the heterogeneous radar network. Spatially distributed radars with different operating at distinct frequencies constitute a heterogeneous radar network, which is modeled as an aircraft recognition graph. In the proposed model, the temporal, spatial and frequency features are extracted and fused in the node, subnet, and global layers, respectively. In node layer, the GRU with the attention mechanism is designed to extract temporal features at each node, and then the spatial features are extracted by the GCN module and then fused by the ATT-GRU module. In the global layer, the temporal-spatial-frequency information is fused and semantic features are distilled for aircraft classification. Extensive experimental results were presented to demonstrate that the proposed model can achieve notable improvements on both reliability and accuracy. Besides, ablation experiment results were carried out to show the expansion from the temporal domain to the temporal-spatial-frequency domain is of great significance for the improvement of reliability and accuracy, which further demonstrates the effectiveness of the proposed method in multi-domain feature extraction and fusion.

    %\section*{Acknowledgment}

    %\bibliographystyle{IEEEtran}
    %\bibliography{ref}
    \bibliographystyle{IEEEtran}
    \bibliography{IEEEabrv,ref1}
\end{document}